\documentclass[12pt]{iopart}
 \usepackage{epsfig}
\begin{document}

\title[Arrays of Cooper Pair Boxes: `Superradiance' and `Revival.']{Arrays of Cooper Pair Boxes Coupled to a Superconducting Reservoir:\\ `Superradiance' and `Revival.'}

\author{D. A. Rodrigues\dag\footnote[3]{Current address: Department of Physics and Astronomy,
University of Nottingham, University Park, NG7 2RD, UK,
(denzil.rodrigues@nottingham.ac.uk)}, B. L. Gy\"orffy\dag\ and T.
P. Spiller\ddag}

\address{\dag H. H. Wills Physics Laboratory, University of Bristol, Tyndall Ave,
 BS8 1TL, UK}

\address{\ddag Hewlett-Packard Laboratories, Filton Road, Bristol, BS34 8QZ,
UK}

\begin{abstract}
We consider an array of $l_b$ Cooper Pair Boxes, each of which is
coupled to a superconducting reservoir by a capacitive tunnel
junction. We discuss two effects that probe not just the quantum
nature of the islands, but also of the superconducting reservoir
coupled to them. These are analogues to the well-known quantum
optical effects `superradiance,' and `revival.' When revival is
extended to multiple systems, we find that `entanglement revival'
can also be observed.
 In order to study the above effects, we
utilise a highly simplified model for these systems in which all
the single-electron energy eigenvalues are set to be the same (the
strong coupling limit), as are the charging energies of the Cooper
Pair Boxes, allowing the whole system to be represented by two
coupled quantum spins, one finite, which represents the array of
boxes, and one representing the reservoir, which we consider in
the limit of infinite size. Although this simplification is
drastic, the model retains the main features necessary to capture
the phenomena of interest.
 Given the progress in superconducting box experiments over recent years,
 it is possible that experiments to investigate both of these interesting
 quantum coherent phenomena could be performed in the forseeable future.
\end{abstract}





\maketitle

\section{Introduction}

Recently there has been much interest, both theoretical
\cite{kato, shnirmanreview} and experimental \cite{nak, saclay,
nak2, delsing}, in superconducting islands coupled to
superconducting reservoirs through Josephson tunnel barriers. Such
structures can be manufactured with energy levels such that the
islands approximate two - level systems, and may have applications
as qubits \cite{shnirman}. In particular, work has focussed on
demonstrating the quantum mechanical nature of these systems
relevant to quantum computing. A full demonstration of many-qubit
quantum algorithms in these systems may be difficult at present,
and it may be helpful to find more general indicators of the
quantum coherent nature of these systems without the need to, for
example, measure a Bell inequality. An ability to demonstrate
coherence is a prerequisite for any quantum computer, and so could
function as a helpful first step. In particular, revival
demonstrates coherence on a timescale which is long compared to
the simple oscillation period. This demonstrates that the
decoherence time is long enough for multiple gate operations to be
performed.

\par In this paper, we discuss two effects that probe not just the quantum nature of the islands, but also of the
superconducting reservoir coupled to them. The first is an
analogue to superradiance in quantum optics \cite{barnett}, in
which the intensity of the radiation emitted from a collection of
atoms is proportional to the square of the number of atoms rather
than linearly proportional. Arrays of Josephson junctions coupled
through a cavity \cite{beasley} have also demonstrated enhanced
emission of radiation (Barbara \etal \cite{lobb}). Here we
consider the current emitted from an array of Cooper Pair Boxes
coupled through a superconducting reservoir, which we call the
super-Josephson effect.

 The second effect is the
phenomenon known in quantum optics as revival. In this, the
coherent superposition properties of an initial quantum state
decay through coupling to the quantum degrees of freedom of the
reservoir, and then `revive' at a later time.

 When revival is extended to two or more two-level systems we discovered a new phenomenon: `entanglement revival.'
\par In order to study the above effects, we utilise a highly simplified model for
the corresponding systems of superconductors. Although the
principle simplification, which is to set all the single-electron
energy eigenvalues to be the same, is drastic, the model retains
the main physical features necessary to capture the phenomena of
interest. For example, the essential feature needed to observe
revival is the discreteness of the Cooper Pairs in the small
superconductors and the reservoir.



\section{The Model}
\subsection{Motivation} \label{sec:motiv}

\par The Josephson effect is an intrinsically quantum mechanical phenomenon,
 as is superconductivity itself. However, in the BCS approximation for a bulk
 superconductor the phase of the order parameter is usually treated as a classical
 variable \cite{BCS}, in the sense that fluctuations about it are considered negligible.
  When the size of the superconductor is reduced, these fluctuations become relevant,
  and the mean-field treatment is no longer valid. Such deviations from mean field
  have
  been observed in experiments on superconducting nanoparticles by Ralph, Black and Tinkham (RBT)
   \cite{RBT}. Subsequent work in the field is well reviewed in \cite{delftsmallreview}.

In what follows we shall be concerned with the same fluctuations
about the mean field theory, albeit for slightly larger
superconducting samples, $\sim 100 nm$, than in the RBT experiment
\cite{RBT}.

\par
In general, to go beyond the BCS approximation is quite difficult
even for a single bulk superconductor. Although an exact solution
exists for a finite size sample \cite{richardson, delftricreview},
this returns a set of coupled equations which rapidly becomes
intractable for a superconductor with more than a few levels.

In our case the difficulty will be compounded by the fact that we
wish to treat an array of small superconductors individually
coupled to a large one. Thus, to render the problem tractable we
shall introduce a highly simplified description of each component
of our system. Nevertheless, we shall endeavour to retain
sufficient realism to capture some generic features of a number of
surprising novel phenomena. These arise when we study the
collective quantum states of an array of small superconductors
which are not coupled to each other but individually coupled to a
common large superconductor; a Cooper Pair reservoir. The model
will be introduced in subsections \ref{sec:CPB} and \ref{sec:res}.
The two new phenomena whose description is the principle aim of
this paper, namely the superconducting analogues of the quantum
optical phenomena of `superradiance' and `revival,' will be
treated in sections \ref{sec:SJ} and \ref{sec:JCrev} respectively.

\subsection{The Cooper Pair Boxes} \label{sec:CPB}

\begin{figure}
\noindent
\epsfig{figure=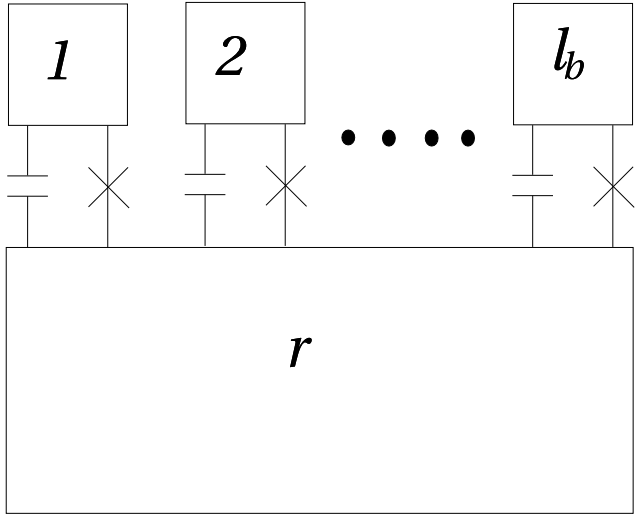,width=8cm,angle=-90} \caption{An array
of $l_b$ superconducting islands, or Cooper Pair Boxes, that are
individually coupled to a superconducting reservoir, $r$, through
capacitive Josephson tunnel junctions (with no inter-box
coupling).} \label{fig:revdiagram}
\end{figure}


The system we shall study is depicted, schematically, in Figure
\ref{fig:revdiagram}. It consists of an uncoupled array of small
superconductors, which we shall refer to as Cooper Pair Boxes,
each of which is coupled through tunnel junctions to the same bulk
superconductor which we shall call the reservoir. The number of
Cooper Pair Boxes will be denoted by $l_b$.


The hamiltonian of the system is:


\begin{equation}
H = H_b+H_r+H_T
\end{equation}

\noindent where the individual hamiltonians refer to the array of
Cooper Pair boxes, the reservoir, and the tunneling between the
two respectively. The first term is the hamiltonian of the boxes:
\begin{equation}
H_b=\sum\limits_{i=0}^{l_b}{E_i^{ch}\sigma^z_i}
\end{equation}

Each Cooper Pair Box is capacitively gated in such a way that it
can be described as a two-level system where the two states are
zero or one excess Cooper Pair on the box, with $E_i^{ch}$ the
charging energy of each. In short, they are the objects
investigated in the experiments of Nakamura \textit{et. al.}
\cite{nak,nak2}.

 The second term $H_r$ is the well known BCS hamiltonian \cite{BCS} (see section \ref{sec:res}), and the final term:

\begin{equation}
H_T=-\sum\limits_{k,i}{T_{i,k}\;( \sigma_i^+ c_{-k
\downarrow}c_{k\uparrow} + \sigma_i^- c_{k\uparrow}^\dag
c_{-k\downarrow}^\dag)}
 \end{equation}

\noindent describes the tunnelling of Cooper Pairs from each of
the boxes to the reservoir (this form can be derived from
single-electron tunnelling \cite{wallace} with $T_{i,k}$
proportional to the square of the single electron tunnelling
energy, $ t_{i,k}^2$). The charging energy of each box can
effectively be tuned by applying a gate voltage to that box, and a
tuneable tunnelling energy can be achieved by adjusting the flux
through a system of two tunnel junctions in parallel
\cite{shnirmanreview}.
 If the charging energies $E_i^{ch}$  and tunnelling elements $T_{i,k}$ of each box are
 all adjusted to be the same, then the charging energy for all the boxes acts like the
 z-component of a quantum spin with $J=l_b/2$. Similarly the sum
 over all the raising (lowering) operators acts like a large-spin
 raising (lowering) operator.

\begin{eqnarray}
 \sum\limits_{i=0}^{l_b} \sigma_i^Z &\rightarrow& S^Z_b \nonumber \\
 \sum\limits_{i=0}^{l_b} \sigma_i^{\pm} &\rightarrow& S^{\pm}_b
\end{eqnarray}

If we consider transitions between the states of the whole system
of boxes caused by only these operators, then the symmetrised
number states form a complete set. The eigenstates of $S^Z_b$
represent states where a given number of the boxes are occupied,
with $m=N_b-l_b/2$. We now have a description of the Cooper-Pair
boxes in terms of a single quantum spin:
\begin{eqnarray}
H_b &=& E^{ch}_bS^Z_b \nonumber\\
H_T &=& \sum\limits_k -T_k (S^+_b c_{-k \downarrow}c_{k\uparrow} +
S^-_b c_{k\uparrow}^\dag c_{-k\downarrow}^\dag)
\end{eqnarray}

\subsection{The Cooper Pair Reservoir} \label{sec:res}

\par The quantum properties of the reservoir will also be of interest and hence we must simplify its description as well. To do this we start with the BCS Hamiltonian written in terms of the Nambu spins \cite{nambu}, $(\sigma^Z_k = \frac{1}{2}(c^{\dag}_{k\uparrow}c_{k\uparrow}+c^{\dag}_{-k\downarrow}c_{-k\downarrow}-1),\; \sigma^+_k=c^{\dag}_{k\uparrow}c^{\dag}_{-k\downarrow}  )$:

\begin{eqnarray} \label{eq:Hr1}
 H_r &=&
  \sum\limits_{k}{2(\epsilon_k - \mu)\sigma_k^Z}-
    \sum\limits_{k=k_F-\omega_c}^{k_F+\omega_c}{(\sigma_k^+\Delta+\sigma_k^-\Delta^*)} \nonumber \\
 \end{eqnarray}

\noindent where $k$ is a generic label (not necessarily the
wavevector) of the single-electron eigenstates and the pairing
field, for an electron-electron coupling constant $V$, is given by
$\Delta_k = V \langle \sigma_k^- \rangle= V \langle
c_{-k\downarrow}c_{k\uparrow}\rangle$. As usual, this has to be
determined self-consistently.

 We shall now make the approximation that the pairing fields, $\Delta_k, \Delta_k^*$, are independent of $k$, and that the term $2(\epsilon_k-\mu)$ is also the same for all $k$. Remembering that the interaction only occurs in a small region around the Fermi energy, $\epsilon_F \pm \omega_c$, the approximation that all the $\epsilon_k$ are equal is equivalent to taking the limit $\omega_c \ll V$ (see \ref{app:intgap}). 
Clearly this is a drastic, strong coupling, approximation
\cite{roman, emile,thouless}, but we will demonstrate that it
nevertheless retains some important features of the
superconducting state. The gain in making this approximation is
that now the Nambu spins can be treated as a single large spin.
Further discussion of the justification of this model when the
superconductor is finite and its relation to the Richardson
solution can be found in \cite{emile}. Note however, that
Yuzbashyan \textit{et. al.} \cite{emile} are concerned only with
isolated islands with a constant number of Cooper Pairs and
therefore discard the diagonal term $S^Z_r$ ($K^Z$).


This representation of a sum over Nambu spins as a single spin was
used by Lee and Scully in 1971 \cite{leeandscully}, but that paper
discarded the commutator $[S^+_{r,\hat{n}} ,
S^-_{r,\hat{n}}]=2S^Z_{r,\hat{n}}$. By retaining it, we can deal
with the discreteness of the Cooper Pairs. The size of the spin is
given by $S_r = l_r/2$ where $l_r$ is the number of levels within
the cutoff region, which is proportional to the volume and will be
taken to infinity at the end of the calculation.
\par In the above limit the hamiltonian (\ref{eq:Hr1}) becomes:

 \begin{eqnarray} \label{eq:spinH}
 H_r &=&
2\xi_{Fr}S^Z_r
    -S^+_r\Delta - S_r^-\Delta^* \nonumber \\
 \end{eqnarray}

\noindent Where $(\epsilon_{Fr}-\mu)$ is written as $\xi_{Fr}$ for
convenience.

Before treating the coupled system of boxes plus reservoir, we
investigate the behaviour of the reservoir alone, in order to
demonstrate that our description still makes sense even after the
above approximation.
\par The hamiltonian of the reservoir (\ref{eq:spinH}) is composed of the $S^Z_r$, $S^+_r$ and $S^-_r$ operators, and so can be written as the component of spin along some other direction $\hat{n}$.

Thus we attempt to diagonalise $H_r$ by defining $S^Z_{r,\hat{n}}
$ by $\gamma_r S^Z_{r,\hat{n}} = H_r$. Then the commutation
relations $[S^Z_{r,\hat{n}} , S^+_{r,\hat{n}}]=S^+_{r,\hat{n}}$
and $[S^+_{r,\hat{n}} , S^-_{r,\hat{n}}]=2S^Z_{r,\hat{n}}$   lead
us to the raising and lowering operators in this direction:

\begin{eqnarray} \label{eq:def-def}
\gamma_r S^Z_{r,\hat{n}} &=& 2\xi_{Fr}S^Z_r -S^+_r\Delta - S_r^-\Delta^*  \nonumber \\
S^+_{r,\hat{n}}&=& dS^Z_{r}+ eS^+_{r} + fS^-_{r} \nonumber \\
|d| &=&|\Delta|/E_F \nonumber \\
e &=& \frac{-d\Delta}{2\xi_{Fr} -2E_F} \nonumber \\
f &=& \frac{-d\Delta^*}{2\xi_{Fr} +2E_F} \nonumber \\
\end{eqnarray}

\noindent Where $\gamma_r=E_F=\sqrt{\xi_{Fr}^2 + |\Delta|^2}$,
i.e. the quasiparticle energy evaluated at the Fermi energy.
\par
The ground state of the hamiltonian is, obviously, the $m=-l_r/2$
eigenstate in the new basis. In the old basis, this is the spin
coherent state \cite{coherent}:

\begin{eqnarray}
|\alpha\rangle &=&
\frac{1}{\sqrt{1+|\alpha|^2}}\sum\limits_{N_r=0}^{l_r}
 \frac{(\alpha^*S_r^+ )^N_r}{N_r!} | 0\rangle
\end{eqnarray}

\noindent which can be rewritten as:

\begin{eqnarray}
|\alpha\rangle &=&
\frac{1}{\sqrt{1+|\alpha|^2}}\prod\limits_{k}{(1 + \alpha^*
\sigma^+_k)} |0\rangle
\end{eqnarray}

As can be readily ascertained using $S_{r,\hat{n}}^Z
|\alpha\rangle =-l_r/2 \; |\alpha \rangle$, choosing $\alpha =
\frac{\xi_{Fr} - E_F}{\Delta}$ ensures that this is the ground
state.
\par
In the second expression for $|\alpha \rangle$ we have explicitly
rewritten the coherent state as a product over $k$. In this form
it is clear that this state is identical to the BCS state in the
limit $\omega_c \ll V$, when $u_k=v_k=1\sqrt{2}$. The pairing
parameter $\Delta$ and the chemical potential $\mu$ are determined
self consistently. In the conventional BCS theory \cite{BCS}:

\begin{eqnarray} \label{eq:BCSconsis1}
\frac{\Delta}{V} &=& \frac{1}{2}\sum\limits_k \frac{\Delta}{E_k} \nonumber \\
\bar{N}_r &=& \frac{1}{2} \sum\limits_k \left( 1 -
\frac{\xi_k}{E_k} \right)
\end{eqnarray}

For our simplified spin model, the fact that the $\epsilon_k$ are
equal allows us to evaluate the sums in (\ref{eq:BCSconsis1}),
giving two coupled algebraic relations instead of integral
equations. Thus we can evaluate $\Delta$ and $\mu$ exactly. We
find:

\begin{eqnarray}\label{eq:spinNdel1}
|\Delta|^2 &=& V^2\bar{N}_r(l_r-\bar{N}_r) \\
\epsilon_{Fr} - \mu &=& V (l_r/2 -\bar{N}_r)  \label{eq:spinNdel2}
\end{eqnarray}

In order to check the validity of these relations, we solve the
gap equations for a general BCS case, and \textit{then} take the
limit where all the levels have the same energy. To take this
limit, we consider a top-hat density of states centred around
$\epsilon_F$ with width $w$ and height $l/w$. This allows the
limit where the bandwidth $w \rightarrow 0$ to be taken whilst
keeping a constant number of levels, $l$. As shown in
\ref{app:intgap}, at zero temperature the integrals can be
calculated exactly. We find:

\begin{eqnarray}
\Delta^2 &=&(\xi_{Fr})^2  \left( \frac{(x+1)^2}{(x-1)^2}-1 \right)
+ \left( \frac{w}{2} \right)^2 \left(
\frac{(x-1)^2}{(x+1)^2}-1 \right) \label{eq:delfromgap}\\
\bar{N}_r &=& \frac{l_r}{w} \left(\frac{w}{2} -
\sqrt{(\xi_{Fr}+\frac{w}{2})^2+\Delta^2} +
\sqrt{(\xi_{Fr}-\frac{w}{2})^2+\Delta^2} \right)
\end{eqnarray}

With $x= e^{\frac{2w}{Vl_r}}$ and $n =\frac{ \bar{N}_r}{l_r}$.
These two equations can be solved to give $\Delta$ and $\xi_{Fr}$.
When the limit $V/w \rightarrow \infty$ (the strong coupling
limit) is taken,  we find that we obtain the same results as given
by the spin model, (\ref{eq:spinNdel1}, \ref{eq:spinNdel2}).
\par
Evidently, the fact that this spin representation reproduces the
BCS results, albeit only in the strong coupling limit, lends
credibility to our simplified model and allows us to consider the
situation we wish to describe, namely that of a series of Cooper
Pair Boxes coupled individually to a single superconducting
reservoir. The problem is then that of two quantum spins, one
finite, which represents the array of Cooper Pair Boxes (labelled
$b$) and one for the reservoir (labelled $r$) which will consider
in the limit $l_r \rightarrow \infty$ at the end of the
calculation. In short, we shall consider the hamiltonian:

 \begin{eqnarray}
 \widehat{H} &=& E^{ch}_bS^Z_b \nonumber\\
 &+& 2\xi_{Fr}S^Z_r
    -S^+_r\Delta - S_r^-\Delta^* \nonumber \\
&-& T (S^+_bS^-_r+S^+_rS^-_b)
 \end{eqnarray}

\section{The Super - Josephson Effect} \label{sec:SJ}
The first effect we consider is an analogy of superradiance (see
e.g. \cite{barnett}). In quantum optics, this refers to the fact
that $l_b$ atoms, placed in a particular state and coupled to a
mode of the electromagnetic field will emit radiation proportional
to the square of the number of atoms. The initial state is one of
the form:
 \begin{equation}
|\psi(0) \rangle= \left\{ |000..111..\rangle  + |001..110..\rangle
+ \cdots + |111..000\rangle \right\}
\end{equation}

\noindent where $0$ and $1$ indicate that an atom (identified by
the place of these numbers in the array which specifies the state
$| n_1,n_2,n_3\dots \rangle$) is in its ground or excited state.
Clearly, this state is an eigenstate of the operator for the total
number of excited states ($S^Z_b$), but not of the individual
$\sigma^Z_i$ operators. Note that it is the symmetrised state with
a total number of excited states $N_b$. Evidently, these are
exactly the states described by the model in section
\ref{sec:CPB}. It is also interesting from the quantum information
theoretic point of view that these are highly entangled states.
These states could, in principle, be generated by tuneable
inter-box coupling \cite{shnirmanreview}. For a carefull
discussion of superradiance the reader is referred to the very
readable account of Eberly \cite{eberly1}. Note also that
Josephson junction arrays placed electromagnetic cavities can emit
radiation with similar $l_b^2$ dependence \cite{lobb}.
\par In this paper, rather than electromagnetic
radiation emitted from atoms (or Josephson junctions) coupled
through a cavity, we consider the current that is emitted from an
array of Cooper Pair Boxes coupled through a common
superconducting reservoir. We consider three cases.
 The first is when the system is in an eigenstate of both the number operator of
 the large superconductor, and the operator measuring the total number of Cooper Pairs on
 the boxes. In this case, whilst our model incorporates the fact
that the electrons are paired, there is no phase coherence between
these pairs.

This is an artificial case if we think of the boxes coupled to a
bulk superconductor,
 but instead would represent many boxes coupled to an island with only a few
levels. If both the box and the reservoir are finite then, of
course, the order parameter $|\Delta|$ is zero. The \emph{square}
of the order parameter $|\Delta|^2$, however, is non-zero,
indicating the presence of pairing between the electrons.
  It is, however, the closest to superradiance, where the atoms are in a symmetrised number
   eigenstate, and the field contains a given number of photons. We
   discuss it here for illustrative purposes.
The second case is when the boxes are in a number eigenstate and
the reservoir in a coherent state, and finally we consider when
the boxes are in a coherent state as well as the reservoir.


\subsection{Superradiant Tunnelling Between Number Eigenstates} \label{sec:N-N}


The simplest case is when both the boxes and the reservoir are in
eigenstates of the number operators $S^Z_b,S^Z_r$ respectively.
The unperturbed hamiltonian is:

\begin{equation}
H = E^{ch}_b S^Z_b + 2\xi_{Fr}S^Z_r
\end{equation}

The eigenstates of this hamiltonian are product states of the
number states $|N_r\rangle$ and $|N_b\rangle$. These states are
equivalent to the usual quantum spin eigenstates $|J,m\rangle$
with $J=l/2$ and $m=N-l/2$.
\par
This hamiltonian is perturbed by the tunnelling hamiltonian:

\begin{equation}
H_T = -T ( S^+_bS^-_r +  S^-_bS^+_r)
\end{equation}

\noindent and the current, i.e. the rate of change of charge on
the island, is given by the commutator of the number operator with
the hamiltonian:

\begin{eqnarray}
\frac{d \hat{N}_b}{dt}=\frac{d
\hat{S}_b^Z}{dt}&=&\frac{1}{i\hbar}[\hat{S}_b^Z, H] \nonumber \\
\hat{I}&=&T ( S^+_bS^-_r -  S^-_bS^+_r)
\end{eqnarray}

The expectation value of the current operator with the ground
state is equal to zero, so we go to time dependent perturbation
theory:

\begin{eqnarray}
|\psi(t)\rangle &=& U(t) |\psi(0)\rangle  \nonumber \\
|\psi(t)\rangle &=& \left( 1 + \frac{1}{i \hbar}
\int\limits_0^t{e^{-i H t'/ \hbar}H_T}e^{-i H t'/ \hbar} {\rm d}t'
+ \cdots \right) |\psi(0)\rangle
\end{eqnarray}

The expectation value of any operator at a given time is, to first
order in $H_T$:

\begin{eqnarray} \label{eq:Ioft}
\langle\psi(t)|\hat{I}|\psi(t)\rangle &=& \langle\psi(0)|\hat{I}|\psi(0)\rangle \nonumber \\
&+& 2 \Re \{ \langle\psi(0)|\hat{I}\frac{1}{i \hbar}
\int\limits_0^t{e^{i H t'/ \hbar}H_Te^{-i H t'/ \hbar} {\rm d}t'}
|\psi(0)\rangle \}
\end{eqnarray}

The tunnelling hamiltonian and current contain only raising and
lowering operators, and so their action on a number eigenstate
produces a simple sum over two other number eigenstates, with a
time dependent factor from the exponentials.



The current is:

\begin{eqnarray}
\langle\psi(t)|\hat{I}|\psi(t)\rangle &=&  \frac{-2T^2}{(i
\hbar)^2} \int\limits_0^t{\cos{ \left((E^{ch}_b -2\xi_{Fr} ) \;t'/
\hbar \right) } {\rm d}t'} \nonumber
\\
 &\times& \left( \langle N_b | S^+_b S^-_b| N_b \rangle \langle
N_r | S^-_r S^+_r| N_r \rangle -\langle N_b | S^-_b S^+_b| N_b
\rangle \langle
N_r | S^+_r S^-_r| N_r \rangle \right) \nonumber \\
\end{eqnarray}

The time dependence is determined by the energy difference between
adjacent number eigenstates, i.e. by $E^{ch}_b $ and $2\xi_{Fr}$.
If we evaluate the expectation values, we find terms like
$N_b(l_b-N_b)(l_r-2N_r)$ which is quadratic in $l_b$ and linear in
$l_r$ (and vice versa, i.e. terms quadratic in $l_r$ and linear in
$l_b$). In particular, to have the largest effect, we set
$N_b=l_b/2$ i.e. half the boxes are occupied.

\begin{eqnarray} \label{eq:eeresult}
\langle\psi(t)|\hat{I}|\psi(t)\rangle &=& \frac{-2T^2}{(i\hbar)^2}
\frac{\sin{ \left((E^{ch}_b -2\xi_{Fr}) t/ \hbar \right) }}{
\left((E^{ch}_b -2\xi_{Fr})/\hbar \right) }
\left( \frac{l_b}{2}(\frac{l_b}{2}+1)(l_r-2N_r) \right) \nonumber \\
\end{eqnarray}

We see that the current is proportional to $l_b^2$, the square of
the number of Cooper Pair boxes. This is in contrast to the
situation where the initial state of the boxes is a product state
of $l_b$ independent wavefunctions, in which any current can be at
most linear in the number of boxes.
 The current is also proportional
to $(n_r-1/2)l_r$, where $n_r = N_r/l_r$, i.e. how far away the
large superconductor is from half filling.


\subsection{Superradiant Tunnelling Between a Number Eigenstate and a Coherent State} \label{sec:N-C}

The second case we consider is when the Cooper Pair boxes are in
an eigenstate of the number operator, and the large superconductor
is in a coherent, that is to say BCS-like, state. We ensure the
coherent state of the large superconductor by introducing a
pairing parameter into the unperturbed hamiltonian:

\begin{equation} \label{eq:HN-C}
H = E^{ch}_b S^Z_b + 2\xi_{Fr} S^Z_r - \Delta_r S_r^+ -\Delta^*_r
S_r^-
\end{equation}

The ground state of the large superconductor is a coherent state
and $\Delta_r = \langle S_r^- \rangle$ can now be found
self-consistently (see section \ref{sec:res}).

We make a basis transform and write the hamiltonian in terms of
this new basis, $\hat{n}$:

\begin{equation}
H = E^{ch}_b  S^Z_b + 2E_{Fr} S^Z_{r,\hat{n}}
\end{equation}

This basis transform makes the unperturbed hamiltonian simpler,
but at the expense of making the perturbation more complicated in
the new basis:

\begin{eqnarray}
H_T &=& -T ( S^+_bS^-_r +  S^-_bS^+_r) \nonumber \\
&=& -T  S^+_b(\frac{2\Delta_r}{E_{Fr} }S^Z_{r,\hat{n}} +
f_r^*S^+_{r,\hat{n}} + e_rS^-_{r,\hat{n}}) \nonumber \\
& & -T  S^-_b(\frac{2\Delta^*_r}{E_{Fr}}S^Z_{r,\hat{n}} +
e_r^*S^+_{r,\hat{n}} + f_rS^-_{r,\hat{n}})
\end{eqnarray}

Again, the tunnelling current is zero to first order in T, as the
boxes are in a number eigenstate. With this basis transform, we
can easily calculate the second-order current. Following the same
procedure as before, we get:

\begin{eqnarray}
\langle\psi(t)|\hat{I}|\psi(t)\rangle &=&
\frac{-2T^2}{(i\hbar)^2}\frac{|\Delta_r|^2}{E_{Fr}^2}\langle
0_{r,\hat{n}}|S^Z_{r,\hat{n}}S^Z_{r,\hat{n}}|0_{r,\hat{n}} \rangle
\int\limits_0^t {\rm d}t' \cos{E^{ch}_b \; t'/\hbar} \nonumber \\
& & \times (\langle N_b | S^+_b S^-_b| N_b \rangle-\langle N_b |
S^-_b S^+_b| N_b \rangle ) \nonumber \\
&-&\frac{2T^2}{(i\hbar)^2}\langle
0_{r,\hat{n}}|S^-_{r,\hat{n}}S^+_{r,\hat{n}}|0_{r,\hat{n}} \rangle
\times
\nonumber \\
& &  \{ |e|^2 \int\limits_0^t {\rm d}t'
\left(\cos{(2E_{Fr}-E^{ch}_b ) \;
t'/\hbar} \right) \langle N_b | S^+_b S^-_b| N_b \rangle \nonumber \\
& & - |f|^2 \int\limits_0^t {\rm d}t' \cos{\left((2E_{Fr}+E^{ch}_b
) \; t'/\hbar \right)} \langle N_b | S^-_b S^+_b| N_b \rangle \}
\end{eqnarray}
The time dependence is again given by the level spacing, but for
the large superconductor this is now given by $E_{Fr} =
\sqrt{\xi_{Fr}^2 + |\Delta|^2}$ rather than $(\epsilon_{Fr}-\mu)$.
\par If we assume the levels of the large superconductor are much more
finely spaced than those of the boxes, i.e. $E_{Fr}\ll E^{ch}_b $,
then:

\begin{eqnarray}
\lefteqn{\langle\psi(t)|\hat{I}|\psi(t)\rangle  }\\
& &=\frac{-2T^2}{(i\hbar)^2} \frac{ \sin{\left(E^{ch}_b t/
\hbar\right)}}{E^{ch}_b / \hbar}\frac{|\Delta_r|^2}{E_{Fr}^2}
\left( \frac{l_r}{2}
\right)^2(l_b-2N_b) \nonumber \\
& &- \frac{2T^2}{(i\hbar)^2} \frac{\sin{ \left(E^{ch}_b t/
\hbar\right)}}{E^{ch}_b / \hbar} l_r \Big\{ \frac{\xi_{Fr}}{E_{Fr}
}N_b(l_b-N_b) +|e|^2N_b -|f|^2(l_b-N_b) \Big\} \nonumber
\end{eqnarray}

Remembering that $l_r \xi_{Fr}/E_{Fr}=  l_r/2-\bar{N}_r $, and
setting $N_b=l_b/2$, we find this is proportional to:

\begin{equation}
l_r(1-2\bar{n}_r)\frac{l_b}{2}(\frac{l_b}{2}+1)
\end{equation}

This is very similar to the case described in (\ref{eq:eeresult})
where both the boxes and the reservoir are in number eigenstates.
In both cases, the effect is largest when the boxes are at
half-filling, and the large superconductor is far from
half-filling. The difference is that here it is the average number
$\bar{N}_r$ that is relevant, not the number eigenvalue.


\subsection{Superradiant Tunnelling Between Coherent States} \label{sec:C-C}

The third case is where both the boxes and the large
superconductors are in the coherent state. Pairing parameters are
introduced for each:

\begin{eqnarray}
H &=& 2\xi_{Fb} S^Z_b - \Delta_b S_b^+ -\Delta^*_b
S_b^- \nonumber \\
&+& 2\xi_{Fr} S^Z_r - \Delta_r S_r^+ -\Delta^*_r S_r^-
\end{eqnarray}

The ground state of this hamiltonian is for both superconductors
to be in a coherent state, which again is determined self
consistently. As before we make a change of basis, this time for
both the boxes and the large superconductor, which makes the
unperturbed hamiltonian:

\begin{equation}
H = 2E_{Fb} S^Z_{b,\hat{n}}+ 2E_{Fr} S^Z_{r,\hat{n}}
\end{equation}

The tunnelling hamiltonian becomes:

\begin{eqnarray}
H_T &=& -T ( S^+_bS^-_r +  S^-_bS^+_r)  \\
&=& -T  (\frac{\Delta^*_b}{E_{Fb} }S^Z_{b,\hat{n}} +
e_b^*S^+_{b,\hat{n}} + f_bS^-_{b,\hat{n}})(\frac{\Delta_r}{E_{Fr}
}S^Z_{r,\hat{n}} +
f_r^*S^+_{r,\hat{n}} + e_rS^-_{r,\hat{n}}) \nonumber \\
& & -T  (\frac{\Delta_b}{E_{Fb} }S^Z_{b,\hat{n}} +
f_b^*S^+_{b,\hat{n}} +
e_bS^-_{b,\hat{n}})(\frac{\Delta^*_r}{E_{Fr} }S^Z_{r,\hat{n}} +
e_r^*S^+_{r,\hat{n}} + f_rS^-_{r,\hat{n}}) \nonumber
\end{eqnarray}

If we insert this into (\ref{eq:Ioft}) we find that we do have a
non-zero term linear in $T$, i.e. the expectation value of
$\hat{I}$ with $|\psi(0)\rangle$. To second order in $T$ we have
terms proportional to $l_r$ and quadratic in $l_b$ and vice versa,
and terms linear in both, i.e.

\begin{eqnarray}
\langle\psi(t) | \hat{I}|\psi(t)\rangle &=&\langle\psi(0) |
\hat{I}|\psi(0)\rangle + \langle\psi(t) |
\hat{I}|\psi(t)\rangle_{l_b^2l_r} \nonumber \\
& &+ \langle\psi(t) | \hat{I}|\psi(t)\rangle_{l_bl_r^2} +
\langle\psi(t) | \hat{I}|\psi(t)\rangle_{l_bl_r}
\end{eqnarray}

The first term is just the expectation value of $\hat{I}$ with the
ground state:

\begin{eqnarray}
\langle\psi(0) | \hat{I}|\psi(0)\rangle
 &=& \frac{T}{i \hbar}
\frac{|\Delta_b|}{2E_{Fb}}\frac{|\Delta_r|}{2E_{Fr}} \;l_b \;l_r\;
2\sin{(\phi_b-\phi_r)}
 \nonumber \\
&=& \frac{T}{i \hbar} \frac{|\Delta_b|}{V_b}\frac{|\Delta_r|}{V_r}
\; 2\sin{(\phi_b-\phi_r)}
\end{eqnarray}

\noindent where $\phi_b, \phi_r$ are the phases of
$\Delta_b,\Delta_r$. This is the usual Josephson effect, as
expected between two BCS-like states. To see how the tunnelling is
enhanced by superradiance, we examine the addtional terms,
\emph{i.e.} the terms quadratic in $l_b$ and $l_r$.
\par

The term quadratic in $l_b$ is:

\begin{eqnarray}
\lefteqn{\langle\psi(t)|\hat{I}|\psi(t)\rangle_{l_b^2l_r}  }\nonumber\\
&=&\frac{-T^2}{(i \hbar)^2} \;l_b^2l_r \;\frac{\sin{2E_{Fr} t/
\hbar}}{2E_{Fr}/ \hbar} \frac{|\Delta_b|^2}{2E_{Fb}^2} \;
\frac{\xi_{Fr}}{E_{Fr}} \nonumber \\
& &\frac{-T^2}{(i \hbar)^2} \;l_b^2l_r \; \frac{\cos{2E_{Fr} t/
\hbar}}{2E_{Fr} / \hbar} \frac{|\Delta_b|^2}{4E_{Fb}^2}
\frac{|\Delta_r|^2}{4E_{Fr}^2} \; 2 \sin{2(\phi_b-\phi_r)}
\end{eqnarray}

\par The first term is proportional to $|\Delta_b|^2/ E_{Fb}^2 $ and
${\xi_{Fr}}/{E_{Fr}}$, and so is largest when the boxes are half
occupied and the large superconductor is far from half occupied.
This term is also phase-independent and is like the effects seen
in sections \ref{sec:N-N} and \ref{sec:N-C}
\par The second term is largest when both are half
occupied, and also depends on the phase difference of the two
pairing parameters with a frequency twice that of the usual
Josephson effect. As such, it can be considered a higher harmonic
in the Josephson current. The term quadratic in $l_r$ is identical
apart from an anti-symmetric exchange of the labels, $b,r$.
\par Finally, the term linear in
both contains a phase - independent term that is largest when both
the boxes and the reservoir are far from half-filling, and a term
that is phase dependent with twice the Josephson frequency that is
largest at half filling.

\begin{eqnarray}
\lefteqn{\langle\psi(t)|\hat{I}|\psi(t)\rangle_{l_bl_r}  }\\
&=&\frac{-T^2}{(i \hbar)^2} \;l_bl_r \;\frac{\sin{2(E_{Fb}+E_{Fr})
t/ \hbar}}{2(E_{Fb}+E_{Fr}) / \hbar} \left(
\frac{\xi_{Fb}^2}{4E_{Fb}^2} \; \frac{\xi_{Fr}}{E_{Fr}} -
\frac{\xi_{Fr}^2}{4E_{Fr}^2} \;
\frac{\xi_{Fb}}{E_{Fb}} \right)\nonumber \\
&+&\frac{T^2}{(i \hbar)^2} \;l_b \;l_r \;\
\frac{\cos{2(E_{Fb}+E_{Fr}) t/ \hbar}}{2(E_{Fb}+E_{Fr})/ \hbar}
\left( \frac{|\Delta_b|^2}{4
E_{Fb}^2}\frac{|\Delta_r|^2}{4E_{Fr}^2} 2\sin{2(\phi_b-\phi_r)}
\right)\nonumber
\end{eqnarray}

In summary we conclude that tunnelling into or out of a Cooper
Pair reservoir from a coherent ensemble of Cooper Pair Boxes can
lead to a tunnelling current proportional to the square of the
number of `boxes.' Interestingly, an observation of such scaling,
in turn, could be taken as a demonstration of coherence and
entanglement of the the `boxes' as such states are necessary to
produce the phenomena.


\section{A `Superconducting Jaynes-Cummings Model' and Revival} \label{sec:JCrev}
\subsection{The Superconducting Analogue of the Jaynes-Cummings Model}

Another well-known quantum optical phenomena is that of quantum
revival (see e.g. \cite{barnett}), which occurs when an atom is
coupled to a single mode of the electromagnetic field. The initial
state of the atom decays, but reappears at a later time. This
apparent decay and subsequent revival occur because the
environment of the atom is a discrete quantum field. Here we
examine a Josephson system analogue. We consider a hamiltonian
(eq. \ref{eq:HN-C}) discussed in section \ref{sec:N-C}, where the
hamiltonian for the box has a number state as a ground state, and
the reservoir ground state is a BCS-like state. These are the
ground states we would expect for an uncoupled box and reservoir.
The tunnelling term $H_T$ couples these states.

\begin{eqnarray} \label{eq:revH}
H &=& E^{ch}_b S^Z_b + 2\xi_{Fr}S^Z_r  - \Delta S^+_r - \Delta^* S^-_r \nonumber \\
H_T &=& -T ( S^+_bS^-_r +  S^-_bS^+_r)
\end{eqnarray}

As in section \ref{sec:N-C}, we make a basis transform so that the
reservoir is written as a quantum spin in another direction.

\begin{eqnarray}
H &=& E^{ch}_b  S^Z_b + 2E_{Fr} S^Z_{r,\hat{n}}\nonumber \\
 H_T &=& -T
S^+_b(\frac{2\Delta_r}{E_{Fr} }S^Z_{r,\hat{n}} +
f_r^*S^+_{r,\hat{n}} + e_rS^-_{r,\hat{n}}) \nonumber \\
& & -T  S^-_b(\frac{2\Delta^*_r}{E_{Fr}}S^Z_{r,\hat{n}} +
e_r^*S^+_{r,\hat{n}} + f_rS^-_{r,\hat{n}})
\end{eqnarray}

We can write a \textit{general} state of the system in terms of
the number eigenstates:

\begin{equation}
|\psi(t) \rangle = \sum\limits_{N_r, N_b=0}^{l_r,l_b}
a_{N_r,N_b}(t) |N_r\rangle |N_b\rangle
\end{equation}

For a single two level system, $l_b=1$, but we can also do the
calculations for a general number of symmetrised boxes.
We can calculate the time evolution of the system using the
Schr\"odinger equations:

\begin{eqnarray}
\lefteqn{i\hbar \sum\limits_{N_r,N_b} \frac{d \,a_{N_r,N_b}(t)
}{dt}|N_r\rangle|N_b\rangle} \nonumber\\
 &=& \sum\limits_{N_r,N_b}
a_{N_r,N_b}(t)
 \left( E^{ch}_b (N_b-\frac{l_b}{2}) + 2\xi_{Fr}(N_r-\frac{l_r}{2}) \right) |N_r\rangle|N_b\rangle\nonumber \\
&-& \sum\limits_{N_r,N_b} a_{N_r,N_b}(t)\,T
S^+_b(\frac{2\Delta_r}{E_{Fr} }S^Z_{r,\hat{n}} +
f_r^*S^+_{r,\hat{n}} + e_rS^-_{r,\hat{n}}) |N_r\rangle|N_b\rangle\nonumber \\
&-& \sum\limits_{N_r,N_b} a_{N_r,N_b}(t)\,T
S^-_b(\frac{2\Delta^*_r}{E_{Fr}}S^Z_{r,\hat{n}}
+e_r^*S^+_{r,\hat{n}} + f_rS^-_{r,\hat{n}}) |N_r\rangle|N_b\rangle \nonumber \\
\end{eqnarray}

More explicitly, we have a set of coupled differential equations
for the coefficients $a_{N_r,N_b}(t)$:

\begin{eqnarray} \label{eq:big_coupled}
i\hbar \frac{d \,a_{N_r,N_b}(t) }{dt} &=& a_{N_r,N_b}(t)
 \left( E^{ch}_b (N_b-\frac{l_b}{2}) + 2\xi_{Fr}(N_r-\frac{l_r}{2}) \right) \nonumber \\
&-&a_{N_r,N_b+1}(t)\,T \sqrt{(N_b+1)(l_b-N_b)}
\frac{2\Delta^*_r}{E_{Fr}} (N_r -l_r/2) \nonumber
\\
&-&a_{N_r-1,N_b+1}(t)\,T \sqrt{(N_b+1)(l_b-N_b)} e_r^*
\sqrt{N_r(l_r-N_r+1)}
\nonumber\\
&-&a_{N_r+1,N_b+1}(t)\,T \sqrt{(N_b+1)(l_b-N_b)}
f_r\sqrt{(N_r+1)(l_r-N_r)}
\nonumber\\
&-&a_{N_r,N_b-1}(t)\,T \sqrt{N_b(l_b-N_b+1)}
\frac{2\Delta_r}{E_{Fr}} (N_r -l_r/2) \nonumber
\\
&-&a_{N_r-1,N_b-1}(t)\,T \sqrt{N_b(l_b-N_b+1)} f_r^*
\sqrt{N_r(l_r-N_r+1)} \nonumber\\
&-&a_{N_r+1,N_b-1}(t)\,T \sqrt{N_b(l_b-N_b+1)}
e_r\sqrt{(N_r+1)(l_r-N_r)}
\nonumber\\
\end{eqnarray}

These equations are tri-diagonal and can be solved numerically.
For now, to stimulate interest, we consider an approximation that
allows a simple analytical model. \par

The $\Delta$ terms in eq. \ref{eq:revH} mean that the total
particle number $N_r+N_b$ is \emph{not} conserved. We note that
whilst the commutator of the total number operator
$[\hat{N}_r+\hat{N}_b,\hat{H}]= \Delta S^+_r - \Delta^* S_r^-$ is
not zero, its expectation value in the coherent state is. With
this in mind, we approximate and assume no fluctuations in the
total number (although we of course retain fluctuations in
$N_r-N_b$) and discard any terms in eq. \ref{eq:big_coupled} that
change the total number.

With these terms discarded, we have a set of coupled differential
equations
 in which each coefficient $a_{N_r,N_b}(t)$ only couples to $a_{N_r-1,N_b+1}(t)$
 and $a_{N_r+1,N_b-1}(t)$.

\begin{eqnarray}
i\hbar \frac{d \,a_{N_r,N_b}(t) }{dt} &=& a_{N_r,N_b}(t) \left( E^{ch}_b (N_b-\frac{l_b}{2}) + E_{Fr}(N_r-\frac{l_r}{2}) \right) \nonumber \\
&-&a_{N_r+1,N_b-1}(t)\,T \sqrt{N_b(l_b-N_b+1)} \sqrt{(N_r+1)(l_r-N_r)} \nonumber \\
&-&a_{N_r-1,N_b+1}(t)\,T \sqrt{(N_b+1)(l_b-N_b)}\sqrt{N_r(l_r-N_r+1)}  \nonumber \\
\end{eqnarray}

\par These equations can be restated in terms of an eigenvalue problem. Each level $N_r$ has a set of eigenvectors for the boxes. If the boxes and the reservoir are initially in a product state, the probability to be in a given value of $N_b$ (regardless of $N_r$) is:

\begin{equation} \label{eq:rev_probs}
P_{N_b} = \sum\limits_{N_b=0}^{l_b} |a_{N_r}(0)|^2 \left|
\sum\limits_i \langle N_b | \nu_{i, N_r} \rangle e^{-iE_{i, N_r}
t/\hbar} \langle \nu_{i, N_r} | \psi_{N_b}(0)\rangle  \right|^2
\end{equation}

\noindent Where $|\nu_{i, N_r} \rangle$ ($E_{i,N_r}$) are the
eigenvectors (values) of the $N_b$-level system associated with
the reservoir level $|N_r \rangle$, $a_{N_r}(0)$ are the initial
amplitudes of $|N_r \rangle$, and $| \psi_{N_b}(0)\rangle  $ is
the initial state of the boxes.
\par This may not be exactly solvable in general, but can be solved for few-level systems, $l_b$. For example, if we have a single two level system and specify the initial state to be a product state of the box state $|0_b \rangle$ with some state $\sum a_{N_r}(0)|N_r\rangle$ of the reservoir, we have the probabilities that the box and reservoir are in a given state:

\begin{eqnarray}
P_{1,N_r-1}(t) &=& |a_{N_r}(0)|^2 \frac{{T'_{N_r}}^2}{{\Omega_{N_r}}^2} \sin^2(\Omega_{N_r}t/\hbar) \nonumber \\
P_{0,N_r}(t) &=& |a_{N_r}(0)|^2 \left( 1- P_{1,N_r-1}(t) \right)
\end{eqnarray}

With $T'_{N_r}=T\sqrt{N_r(l_r-N_r+1)}$ and $\Omega_{N_r} = \sqrt{
(E^{ch}_b - E_{Fr})^2/4 + {T'}^2}$. This is the same result as in
quantum optics, except that the photon creation/ annihilation
operators give $T'_{N_r}=T\sqrt{N_r}$ .

\subsection{Quantum Revival of the Initial State}

In quantum optics, the phenomenon of revival is seen when the
electromagnetic field is placed in a coherent state (i.e.
$|a_{N}|^2 = \exp(-\bar{N})\bar{N}^N/N!$, with the sum over $N$
running to infinity). We place the reservoir in the \textit{spin}
coherent state, i.e.

\begin{equation}
|a_{N_r}(0)|^2 = |\alpha|^{2N_r}\frac{l_r!}{(l_r-N_r)!N_r!}
\end{equation}

\noindent Where $\alpha$ is determined by the average number
$|\alpha| = \sqrt{\bar{N_r}/(l_r-\bar{N}_r)}$.
\par The probability of the box being in the state $|0_b\rangle $ at a given time is:

\begin{equation}
P_0(t) = \sum\limits_{N_r}^{l_r} |a_{N_r}(0)|^2 P_{0,N_r}(t)
\end{equation}

\begin{figure}
\noindent
\epsfig{figure=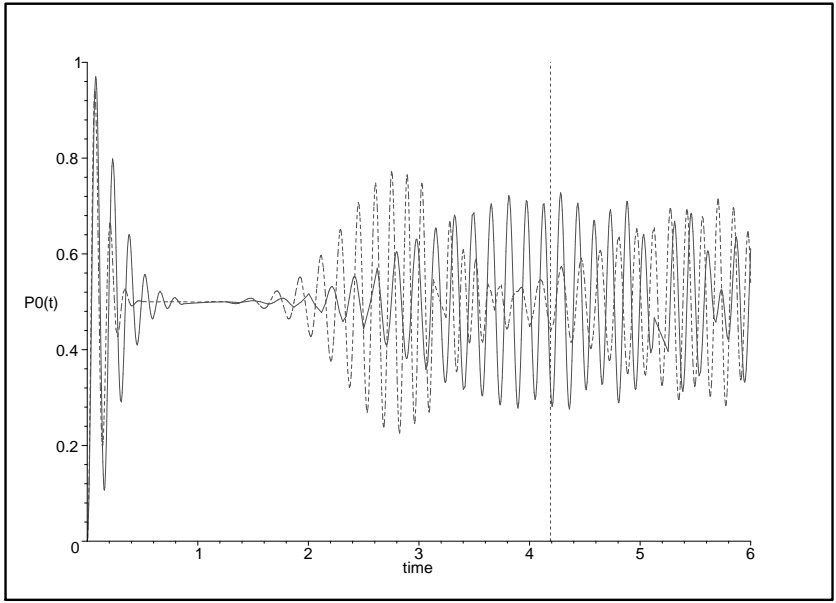,width=8cm,angle=-90} \caption{Spin
Coherent State Revival for $N_r=10$, $l_r=50$ (solid line). The
faint line indicates the usual quantum optical coherent state
revival with $N_r=10$, scaled along the $t$ axis by $\sqrt{l_r}$
and the vertical dash-dot line indicates the calculated revival
time.} \label{fig:rev1}
\end{figure}

For simplicity we consider the case when $E^{ch}_b = E_{Fr}$, i.e.
$\Omega_{N_r} =T'_{N_r}$. This is shown in Figure \ref{fig:rev1},
for the values $\bar{N}_r=10$, $l_r=50$. The initial state dies
away, to be revived at a later time.

\par The revival occurs when the terms in the sum are in phase. We can make an estimate of this by requiring the terms close to $\bar{N}_r$ to be in phase:

\begin{eqnarray}
2\pi &=& 2T\sqrt{(\bar{N}_r+1)(l_r-\bar{N}_r)}\;t_{rev} - 2T\sqrt{\bar{N}_r(l_r-\bar{N}_r+1)} \;t_{rev}\nonumber \\
t_{rev}  &=& \frac{2\pi}{T}
\frac{\bar{n}_r^{1/2}(1-\bar{n}_r)^{1/2}}{(1-2\bar{n}_r)}
\end{eqnarray}

Where, as before $n_r = N_r/l_r$. Note that the revival time
remains finite in the $l_r \rightarrow \infty$ limit, as long as
$N_r/l_r$ is finite.

We can find an analytic form for the initial decay, and also
demonstrate the equivalence of this `spin revival' to the well
known quantum optics revival, in the limit $l_r \rightarrow
\infty$, $l_r/N_r \rightarrow \infty$. Note that this is an
unrealistic limit for our case, where we would keep the filling
factor $N_r/l_r$ constant, but it allows us to make a connection
with the quantum optics case.

\begin{figure}
\epsfig{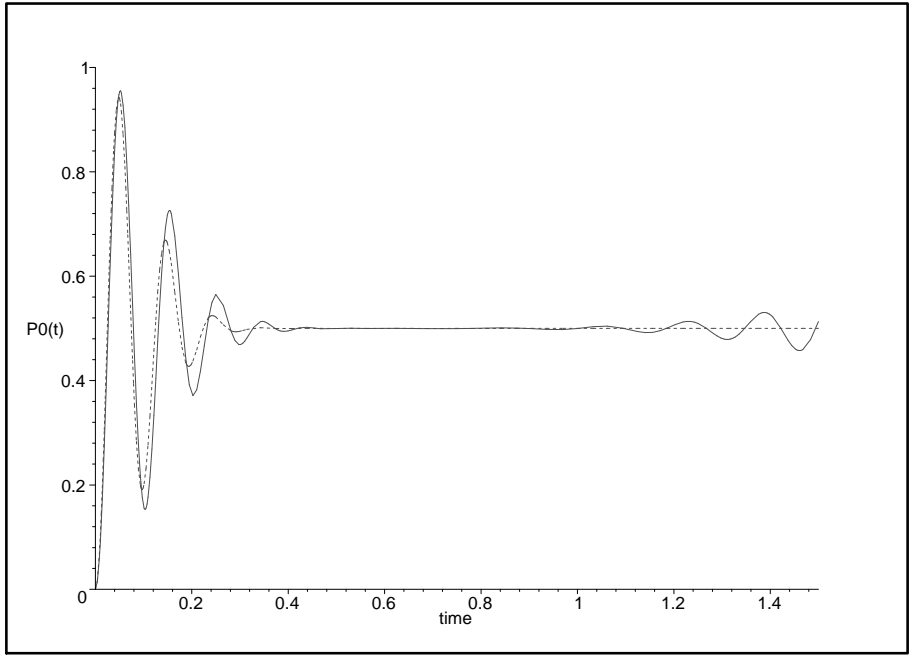} \caption{The
large $\bar{N_r}, l_r$ limit: Solid line indicates single box
revival for  $N_r=10$, $l_r=100$, and the dashed line indicates
the asymptotic form for the decay.} \label{fig:revlimit}
\end{figure}

The coefficients, $a_{N_r}(0)$ for the coherent and spin-coherent
states are the Poisson and binomial distributions respectively,
and both can be approximated by a Gaussian distribution.

\begin{eqnarray}
\exp(-\bar{N}_r)\bar{N}_r^{N_r}/N_r!&\simeq& (2\pi \bar{N}_r)^{-1/2} \exp \left[ -(N_r-\bar{N}_r)^2/2\bar{N}_r \right] \nonumber \\
|\alpha|^{2N_r}\frac{l_r!}{(l_r-N_r)!N_r!} &\simeq& (2\pi
\bar{N}_r)^{-1/2} \exp \left[ -(N_r-\bar{N}_r)^2/2\bar{N}_r
\right]
 \end{eqnarray}

For the coherent state we have taken the limit $ \bar{N}_r \gg 1$.
For the spin-coherent case we have also assumed $l_r \gg
\bar{N}_r$. The gaussian factor suppresses any terms not around
the average $\bar{N}_r$, so we can expand $N_r \simeq \bar{N}_r +
(N_r -  \bar{N}_r)$.
\par In the limit, the spin coherent state gives:

\begin{eqnarray}
P_0(t) &=& \frac{1}{2} + \frac{1}{2(2\pi \bar{N})^{1/2}} \int {\rm d}N \, \exp \left( -\frac{(N_r-\bar{N}_r)^2}{2\bar{N}_r} \right) \nonumber \\
& & \times \cos \left[ 2 T ({l_r \bar{N}_r})^{1/2}t
\left(1+\frac{(N_r-\bar{N}_r)^2}{2\bar{N}_r }\right) \right]
 \end{eqnarray}

This is the same as for the coherent state apart from the factor
of $l_r^{1/2}$ if the frequency. Doing the integral gives:

\begin{equation}
P_0(t) = \frac{1}{2} + \frac{1}{2} \cos(2T (l_r\bar{N}_r)^{1/2}
t)\; \exp(-\frac{(Tl_rt)^2}{2})
\end{equation}

This is plotted in Figure \ref{fig:revlimit}.




Thus the phenomenon of quantum revival has a direct analogy in our
`spin superconductor' model.  In both cases, the revival is due to
constructive interference between the terms in the sum over the
reservoir (field) number states.

\subsection{Revival of Entanglement}

Revival is usually considered for single two-level systems, but
the formula in (\ref{eq:rev_probs}) gives the state for a general
system. In particular, the dynamics of a pair of two-level systems
can be easily calculated, either numerically, or analytically,
remembering that the singlet state is uncoupled due to radiation
trapping.
\par
\begin{figure}
\epsfig{figure=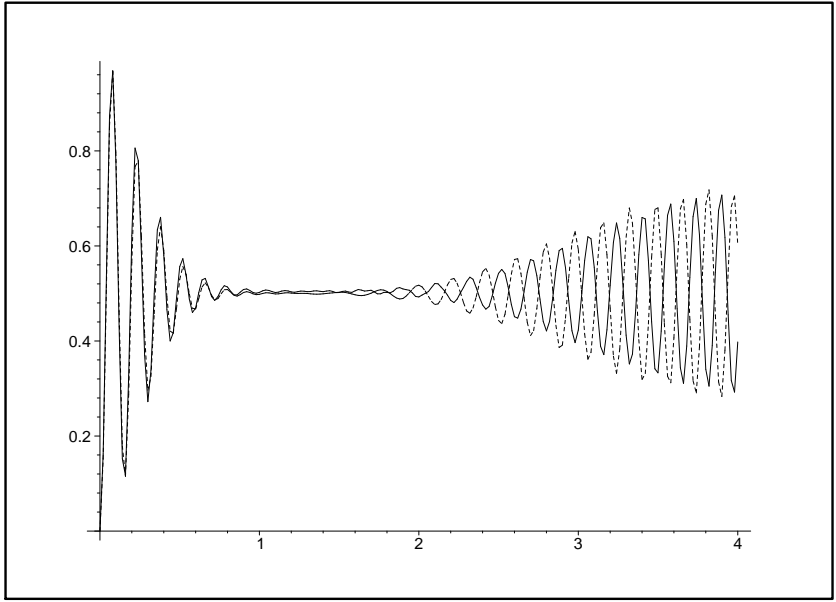,width=8cm,angle=-90} \caption{Two Cooper
Pair Box Spin Coherent State Revival for $N_r=10$, $l_r=50$. The
solid line represents the probability that, when the system is
initialised in the state $|00\rangle$, one of the Cooper pair
boxes is in the state $|0\rangle$ at time $t$, after tracing over
the other box. The dashed line indicates the revival of a single
isolated Cooper Pair box connected to a reservoir. Note that the
revivals occur at the same time, but out of phase.}
\label{fig:2bitrev}
\end{figure}

 Starting in the state $|00\rangle$, we find a very similar effect to the single-island case. When one island is traced out, the revival of the other occurs at the same time as a single island, but with opposite phase (Figure \ref{fig:2bitrev}).

If we do not trace over one island, but consider the revival of
the state $|00\rangle$, we find oscillations in the probability of
returning to this state earlier than in the case of a single
island.
\par
\begin{figure}
\epsfig{figure=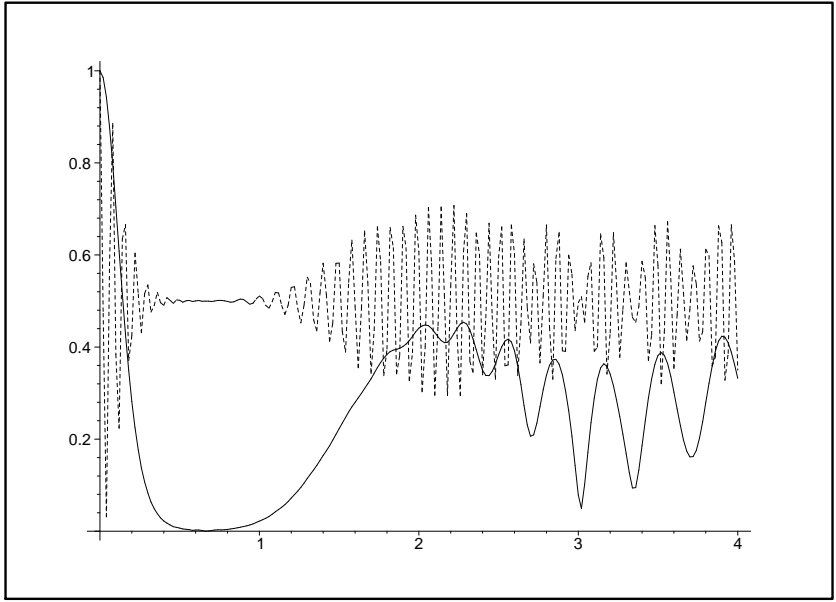,width=8cm,angle=-90} \caption{Two Cooper
Pair Box Spin Coherent State Revival for $N_r=10$, $l_r=50$. The
dashed line represents the probability that the Cooper Pair boxes
have returned to their initial state
$(|01\rangle+|10\rangle)/\sqrt{2}$ at time $t$. The dashed line
indicates the negativity of the boxes as a function of time.}
\label{fig:2bitrevent}
\end{figure}

 To investigate the revival of entanglement, we start the system in the state $(|01\rangle+|10\rangle)/\sqrt{2}$.
As before we note that the oscillations in the probability of
being in the state $(|01\rangle+|10\rangle)/\sqrt{2}$ decay and
then revive. It is interesting to note, however, that in this case
the initial state is an entangled one. One measure of entanglement
is the negativity \cite{neg}, i.e. the sum of the negative
eigenvalues of the partially transposed density matrix. If we plot
the negativity over time, we see that it dies away at first, as do
the oscillations. As the oscillations revive, we also see an
revival in the negativity of the two islands (Figure
\ref{fig:2bitrevent}).

\par This revival of entanglement, an initially counterintuitive phenomenon, is of course due to the fact that we have considered the entanglement between the two islands only. When the entanglement `disappears,' it is only the entanglement between the islands that has disappeared. If we were to consider the whole system, including the reservoir, the evolution would be unitary and the entanglement would remain constant.
\par It is worth noting that this `entanglement revival' is not a unique feature of our spin model, and would be expected whenever two-level systems are coupled to a field with (a possibly infinite number of) discrete energy levels.

\section{Conclusions}

\par We have presented a simple model of a superconductor that allows an
 easy investigation of various phenomena associated with fluctuations around
  the BCS, mean field, behaviour. By taking the limit where all the single-electron
   energy levels are equal, we can represent the superconducting hamiltonian in terms
    of the operators of a large, finite quantum spin, which we then consider in
    the limit $l_r \rightarrow \infty$.
     Comparison of results using this
     model in the mean field with the limit of BCS results gives us confidence
that whilst the model can be regarded as a caricature of a
realistic description it retains some generic features of quantum
fluctuations. Furthermore, we have demonstrated that the
simplicity of the model allows the easy investigation of several
surprising physical phenomena.
\par As the first of these we considered an analogue of the quantum optical
 effect of superradiance, where the emitted intensity of a number of atoms
 coupled to an electromagnetic field is proportional to the \textit{square
 } of the number of atoms.


In the analogous `Super-Josephson effect,' the current from a set
of Cooper Pair boxes connected to a reservoir is proportional to
the \textit{square} of the number of boxes $l_b$. This effect is
due to the entangled nature of the initial state. Any product
state (or mixture of product states) would produce a current at
most linear in $l_b$. The effect also relies on the reservoir
being away from the $N_r/l_r=1/2$ state, i.e. away from
half-filling.
    \par The analogy has been extended to the physically relevant case case where the reservoir
    is in a coherent (\emph{i.e.} BCS-like) state and the boxes are in a number state. As before we have
no current to first order in $T$, but have a current proportional
to $l_b^2$ to second order. Also as in the previous case, the
effect requires the reservoir to be far from half-filling, but
here it is the average number $\bar{N}_r$ that must be far from
$l_r/2$.
\par Finally we considered the effect when both the
reservoir and the boxes are initially in a coherent state. In
addition to the  usual Josephson
 current, we observe the effect described above and also find a current that is dependent on the
 phase difference between the superconductors, with twice the phase dependence of the usual
 Josephson effect.
\par Interestingly, the above discussion suggests that an observation of a Josephson current
 which scales quadratically with the number of small superconductors (qubits) can be taken as
 evidence for their entanglement. As experiments are now at the level of entangling two charge
  qubits \cite{nak2}, and these effects can be detected by making only current measurements,
  the above effects could be usefully looked for experimentally.

\par We have also studied an analogue of quantum revival (again considering the case a box in a number
state coupled to a reservoir in a BCS-like state) , in which the
oscillations (of the type observed in \cite{nak, saclay, nak2,
delsing}) in the probability of occupation of an island decay at
first, only to revive at a later time. This effect can also be
seen when we have two Cooper Pair boxes coupled to a reservoir. If
we place these boxes in an entangled state and calculated the
entanglement over time, we see that the entanglement between the
boxes dies away at first, only to revive later. This remarkable
effect demonstrates and measures the quantum fluctuations of the
reservoir superconductor about the BCS mean field theory. In other
words it probes the discreteness of the Cooper Pairs in the same
way as the corresponding quantum optical effect probes the
discreteness of photons of the radiation field.

\subsection*{Acknowledgements}
We would like to thank James Annett for helpful discussions.
Denzil Rodrigues is supported by an EPSRC CASE studentship with
sponsorship from Hewlett-Packard.


\appendix

\section{Limit of BCS Self Consistency Equations} \label{app:intgap}

As confirmation of the validity of our spin model, we wish to
compare it with the standard BCS theory. In particular, we wish to
check that solving the gap equations from the spin model gives the
same result (\ref{eq:spinNdel1}, \ref{eq:spinNdel2}) as if we
solve the usual BCS gap equations (\ref{eq:BCSconsis1}) first, and
then take the limit. We wish to take the limit where the
single-electron energy levels, $\epsilon_k$ are equal. To do this
we consider a density of states which is a top-hat function
centred around $\epsilon_F$ with width $\omega$ and height
$l_r/\omega$. Taking the limit $\omega \rightarrow 0$ ensures that
the interaction occurs in a narrow region around $\epsilon_F$
whilst ensuring the integral is over a constant number of levels,
$l_r$. We have two integral equations that must be solved
self-consistently. The first is an equation for the pairing
parameter, $\Delta$:

\begin{eqnarray} \label{eq:BCSconsis2}
\frac{\Delta}{V} &=&
\frac{1}{2}\int\limits_{\epsilon_F-\omega}^{\epsilon_F+\omega}
{\rm d}\epsilon \frac{\Delta}{
((\epsilon-\mu)^2+|\Delta|^2))^\frac{1}{2} }
\end{eqnarray}

\noindent and the second is an equation for the average number of
Cooper Pairs, $\bar{N}$:

\begin{equation}
\bar{N} = \frac{1}{2}
\int\limits_{\epsilon_F-\omega}^{\epsilon_F+\omega} {\rm
d}\epsilon \left( 1 -
\frac{\epsilon-\mu}{((\epsilon-\mu)^2+|\Delta|^2))^\frac{1}{2}}
\right)
\end{equation}

As we are at zero temperature, both these integrals can be done
exactly:
\begin{eqnarray} \label{eq:delint1}
\frac{1}{V} &=& \frac{l}{2\omega} \ln \left(  \frac{ (\xi_F+\omega/2)+\sqrt{(\xi_F+\omega/2)^2+|\Delta|^2} }{(\xi_F-\omega/2)+\sqrt{(\xi_F-\omega/2)^2+|\Delta|^2}}      \right) \\
\bar{N} &=& \frac{l}{\omega} \left( \frac{\omega}{2} -
\sqrt{(\xi_F+\omega/2)^2+|\Delta|^2}+\sqrt{(\xi_F-\omega/2)^2+|\Delta|^2}
\right) \label{eq:delint2}
\end{eqnarray}

Again, $\xi_F$ is shorthand for $\epsilon_F-\mu$. These can be
rearranged to give two equations for $|\Delta|^2$
(\ref{eq:deleqns1}, \ref{eq:deleqns2}), or two equations for
$\xi_F^2$ (\ref{eq:xieqns1}, \ref{eq:xieqns2}). The equations for
$|\Delta|^2$ are:

\begin{eqnarray} \label{eq:deleqns1}
|\Delta|^2 &=&(\xi_F)^2  \left( \frac{(x+1)^2}{(x-1)^2}-1 \right)
+ \left( \frac{w}{2} \right)^2 \left(
\frac{(x-1)^2}{(x+1)^2}-1 \right) \\
|\Delta|^2 &=& \frac{\bar{N}(l-\bar{N})}{l^2(l-2\bar{N})^2}
\left(4{\xi_F}^2l^2 -\omega^2(l-2\bar{N})^2 \right)
\label{eq:deleqns2}
\end{eqnarray}

Equating the two and rearranging gives us an expression for
${\xi_F}^2$:

\begin{equation} \label{eq:xieqns3}
{\xi_F}^2 =\left(\frac{\omega}{2} \right)^2 \left(
\frac{(x+1)^2}{(x-1)^2}-1 +4(1-2n)^2   \right) \bigg/ \left(
\frac{4n(1-n)}{(1-2n)^2} -  \frac{(x-1)^2}{(x+1)^2}+1\right)
\end{equation}

\noindent where $x= e^{\frac{2w}{Vl}}$ and $n =\frac{
\bar{N}}{l}$. If we rearrange eqns. (\ref{eq:delint1},
\ref{eq:delint2}) for $\xi_F$ we get:

\begin{eqnarray}\label{eq:xieqns1}
{\xi_F}^2 &=& \left(\frac{\omega}{2} \right)^2
\frac{(x+1)^2}{(x-1)^2} - |\Delta|^2 \frac{(x+1)^2}{4x} \\
{\xi_F}^2 &=& \left(\frac{\omega}{2} \right)^2 (1-2n)^2 +
|\Delta|^2 \frac{(1-2n)^2}{4n(1-n)}\label{eq:xieqns2}
\end{eqnarray}

Which yield:

\begin{eqnarray} \label{eq:deleqns3}
|\Delta|^2 = \left(\frac{\omega}{2} \right)^2 \left(
\frac{(x+1)^2}{(x-1)^2}- (1-2n)^2  \right) \bigg/ \left(
\frac{(1-2n)^2}{4n(1-n)} +\frac{(x+1)^2}{4x}   \right)
\end{eqnarray}

From eqns. (\ref{eq:xieqns3}, \ref{eq:deleqns3}), it is clear that
the correct limit to take is $\omega/V \rightarrow 0$. Taking this
limit of eqns. (\ref{eq:xieqns3}, \ref{eq:deleqns3}), we get:

\begin{eqnarray}
{\xi_F}^2 &=& \frac{V^2 l^2}{4}(1-2n)^2\\
|\Delta|^2 &=& V^2 l^2 n(1-n)
\end{eqnarray}

That is, we have regained the forms for $\Delta$ and $\xi$ given
by the spin model (\ref{eq:spinNdel1}, \ref{eq:spinNdel2}).

\end{document}